\documentclass[preprint2, twocolappendix]{aastex7}
\usepackage{amsmath,amsmath,ulem}

\usepackage{amsmath,amsmath,ulem}
\usepackage{tikz}
\usepackage{ifthen}
\usepackage[flushleft]{threeparttable}
\usepgflibrary{fpu}
\graphicspath{{./}{figures/}}

\tabletypesize{\footnotesize}

\begin{document}

\title{Saturnian Irregular Satellites as a Probe of Kuiper Belt Surface Evolution}

\author[0000-0003-4778-6170]{Matthew Belyakov}
\email{mattbel@caltech.edu}
\affiliation{Division of Geological and Planetary Sciences, California Institute of Technology, Pasadena, CA 91125, USA}
\correspondingauthor{Matthew Belyakov}
\author[0000-0002-8255-0545]{Michael E. Brown}
\email{mbrown@caltech.edu}
\affiliation{Division of Geological and Planetary Sciences, California Institute of Technology, Pasadena, CA 91125, USA}

\begin{abstract}
We present the JWST NIRSpec PRISM 0.7-5.3 $\micron$ spectra of Albiorix and Siarnaq and the NIRSpec G235H/G395M 1.7-5.3 $\micron$ spectra of Phoebe, the three largest Saturnian irregular satellites. The irregular satellites of the giant planets are thought to be captured planetesimals from the same population as Kuiper belt objects. They are emplaced inside Saturn's Hill sphere during the giant-planet instability described by the Nice Model, and are thus valuable tracers of Kuiper belt surface evolution. Phoebe's JWST spectrum matches the global average from Cassini VIMS, and by comparing the spectrum to the library of Kuiper belt object spectra from JWST, we demonstrate Phoebe's compositional similarity to water-rich KBOs. On the smaller Albiorix and Siarnaq, we observe a broad 3 $\micron$ O-H band but do not see a Fresnel peak or the 1.5/2.0 $\micron$ features characteristic of H$_2$O ice. We posit that after capture, the frequent high-velocity collisions between smaller irregular satellites sublimate the water ice, while the much larger Phoebe is resistant to disruption and retains its water ice. We suggest that the presence of CO$_2$ on the smaller satellites, despite the lack of water ice, indicates later formation of CO$_2$ on these surfaces through irradiation of organic compounds.
\end{abstract}

\section{Introduction}

The irregular satellites of the giant planets are a population of small bodies on distant, eccentric, and highly-inclined or retrograde orbits that preclude in-situ formation as a possible origin \citep{1956VA......2.1631K}. Multiple capture mechanisms from initially heliocentric orbits have been proposed in the past, such as capture by drag from the gas surrounding circumplanetary disks \citep{1979Icar...37..587P}, capture from rapid mass accretion by giant planets in their early stages of growth \citep{Heppenheimer1977Icar}, and collisions between asteroids \citep{Colombo1971Icar}. The modern mechanism for irregular satellite capture was proposed by \cite{2007AJ....133.1962N}, who demonstrate that during giant planet migration in the early solar system \citep{Morbidelli2010CRPhy, Nesvorny2018ARA&A}, close encounters between planetary-sized bodies perturb planetesimals that happen to be near the Hill spheres of the giant planets onto planet-bound orbits. Capture through third-body assist reproduces the observed size-frequency and orbital distributions of the irregular satellites \citep{Bottke2010AJ, 2014ApJ...784...22N}. This version of the capture mechanism implies that the irregular satellites share the same source population with the Kuiper belt and the Jupiter Trojans \citep{Morbidelli2010CRPhy}. While three-body capture from binary asteroids as demonstrated in \cite{Agnor2006Natur} could occur from a source region of planetesimals local to the giant planets \citep{Jewitt2007ARA&A}, irregular satellites captured pre-dynamical instability will be lost \citep{2007AJ....133.1962N}.  

While the dynamical link between irregular satellites and Kuiper belt objects (KBOs) is well-established, the compositional connection remains incomplete. KBOs in the size range of irregular satellites have bifurcated visible spectral slopes split into two modes at 10 and 30 \%/100 nm, sometimes termed ``red'' and ``very red''  \citep{1998Natur.392...49T, 2002AJ....124.2279D, 2012A&A...546A.115H,2017AJ....153..145W}. By contrast, the colors of the irregular satellites are more similar to those of the Jupiter Trojans, with typical spectral slopes between 0-15 \%/100 nm  \citep{Grav2007Icar, Maris2007A&A, Graykowski2018AJ, 2022AJ....163..274P}. Spectroscopic studies of the Jupiter irregular satellites have shown them to be distinct from KBOs, with featureless spectra similar, though not identical, to the Jupiter Trojans  \citep{Brown2014ApJL, Sharkey2023PSJ}. Spectroscopy of the Uranian irregular satellites has not found conclusive evidence for surface ices and volatiles \citep{Romon2001A&A, Vilas2006Icar, Sharkey2023PSJ}, though water ice has been detected on Nereid, the largest of the Neptunian irregular satellites \citep{Brown1998ApJL, Brown1999Icar}. 

\begin{deluxetable*}{lcccccc}[t!]
\tablewidth{0pc}
\setlength{\tabcolsep}{8pt}
\tabletypesize{\normalsize}
\renewcommand{\arraystretch}{1.0}
\tablecaption{
    JWST NIRSpec Observations
    \label{tab:obs}
}
\tablehead{\colhead{Target} & \multicolumn{1}{c}{Date and Time} & \multicolumn{1}{c}{$r$ (au)\tablenotemark{\scriptsize 1}} & \multicolumn{1}{c}{$R$ (au)\tablenotemark{\scriptsize 1}}& \multicolumn{1}{c}{$\alpha$ (deg)\tablenotemark{\scriptsize 1}} & \multicolumn{1}{c}{mag \tablenotemark{\scriptsize 1}} & \multicolumn{1}{c}{$t_{\mathrm{exp}}$ (s)\tablenotemark{\scriptsize 2}}  
}
\startdata
Albiorix & 2023 November 22, 01:20-03:00 & 9.86 & 9.79 & 5.81 & 21.1 & 2393 \\
Phoebe & 2023 November 20, 22:00-23:39 & 9.78 & 9.68 & 5.85 & 16.7 & 1109 \\
Siarnaq & 2023 November 22, 03:01-04:05 & 9.78 & 9.73 & 5.85 & 20.5 & 1517 \\
\enddata
\vspace{-0.1cm}\tablenotetext{{1}}{ Heliocentric distance $r$, distance from JWST $R$, phase angle $\alpha$, and $V$-band magnitude of Phoebe at time of observation, ground-based r-mag for Albiorix and Siarnaq from \cite{2022AJ....163..274P}.}
\vspace{-0.15cm}\tablenotetext{{2}}{ Total exposure times, Albiorix and Siarnaq are observed in PRISM mode, while Phoebe is observed in G235H/F170LP and G395M/F290LP, with the same exposure time in both dither settings.}
\vspace{-1cm}
\end{deluxetable*}

JWST NIRSpec has uncovered a wide range of outer solar system surface compositions, showing that KBOs commonly have H$_2$O, CO$_2$, CO, and organic ices on their surfaces \citep{Brown2023PSJ, DePra2024NatAs, Souza-Feliciano2024A&A, Licandro2024NatAs, Pinilla-Alonso2024NatAs}, while the Jupiter Trojans have weak 3 and 3.4 $\micron$ absorptions attributed to hydrated minerals and organics respectively, with a notable single detection of CO$_2$ \citep{Wong2024PSJ}. This enhanced picture of the surfaces of outer solar system small bodies allows us to begin drawing comparisons between KBOs, Centaurs, Trojans, and irregular satellites, which are predicted by dynamical instability models to be part of the same initial source population. In this work we present the JWST NIRSpec 0.7-5.0 $\micron$ spectra of the three largest Saturnian irregular satellites: Phoebe, Siarnaq, and Albiorix. We briefly summarize their known properties.

Phoebe is the largest of the Saturnian irregular satellites with a diameter of 212 km and a retrograde orbit. It is the parent body for the material in its eponymous ring \citep{Verbiscer2009Natur}, whose particles darken Iapetus's leading side. The 2004 \textit{Cassini} fly-by estimated Phoebe's density to be 1.6 g cm$^{-3}$, which combined with a low albedo of 0.1 \citep{Grav2015ApJ} indicate capture from the primordial planetessimal disk \citep{Porco2004SSRv, Johnson2005Natur}. Spectral imaging of the surface found varying amounts of spatially anti-correlated water ice and CO$_2$ \citep{Hansen2012Icar}, along with other potential absorption features \citep{Clark2005Natur, Coradini2008Icar}. Later work by \cite{Fraser2018AJ} confirmed that water ice is ubiquitous on Phoebe's surface and is richest near impact basins, suggesting collisions are exposing subsurface ice. Prior to JWST, there did not exist KBO spectra with similar wavelength coverage as those of Phoebe, but photometrically, Phoebe's color is distinctly blue (g-i = 0.33) as opposed to the red KBOs \citep{Grav2007Icar,2022AJ....163..274P}. However, colors of its different surface types span the full extent of the KBO color range from solar-like to very red \citep{Fraser2018AJ}.

Siarnaq (39 km) and Albiorix (29 km) are the largest members of the prograde Gallic and Inuit groups of Saturnian irregular satellites, respectively \citep{2015ApJ...809....3G, Denk2018eims}. Both have darker albedos than Phoebe, at 5-6\%, but have significantly redder colors (g-i = $0.77$), similar to those of the Jupiter Trojans \citep{Grav2007Icar, Grav2015ApJ, Graykowski2018AJ, 2022AJ....163..274P}. 

We begin by presenting the reduction of the JWST NIRSpec observations of Phoebe, Siarnaq, and Albiorix in \autoref{sec:obs}. In \autoref{sec:spec} we show three key spectral features on the Saturnian irregular satellites in the context of available JWST Kuiper belt and Centaur data: H$_2$O, organics, and CO$_2$. Finally, in \autoref{sec:discussion} we discuss the possible reasons for the differences between Phoebe and the smaller Albiorix and Siarnaq, particularly in the lack of water ice on the smaller satellites.

\begin{figure*}
    \centering
    \includegraphics[width=\textwidth]{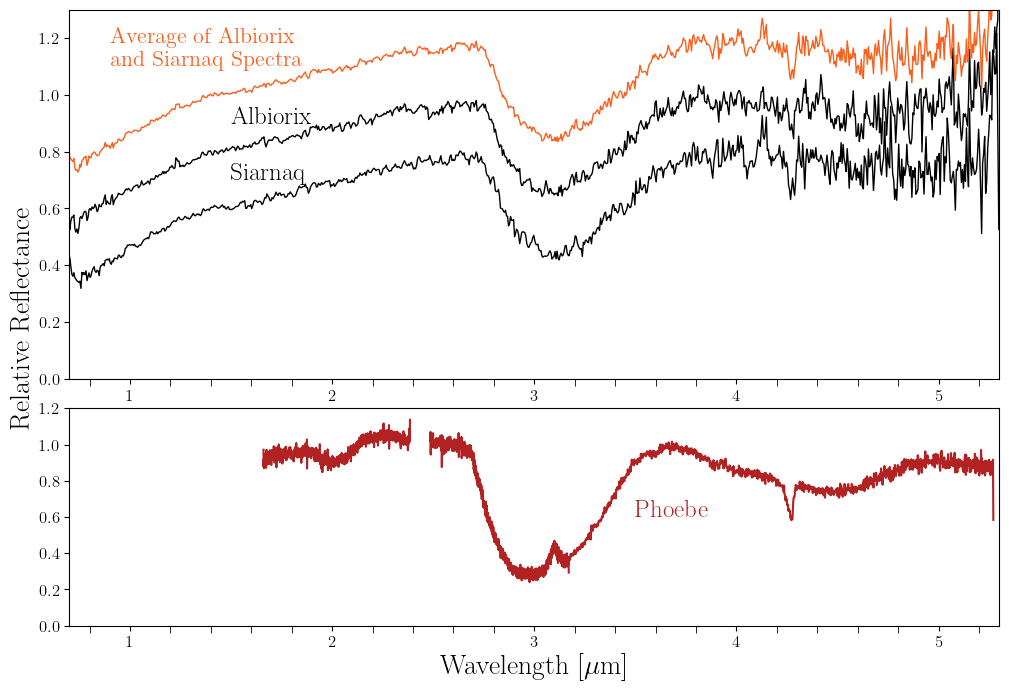}
    \caption{Relative reflectance spectra of the three largest Saturnian irregular satellites, Siarnaq, Albiorix, and Phoebe. Top panel shows the individual spectra of Albiorix and Siarnaq and their averaged spectrum, with Albiorix and Siarnaq offset down by 0.2 and 0.4 respectively. These two smaller irregular satellites show a broad 3 $\micron$ absorption, a weak organic feature at 3.4 $\micron$, and CO$_2$ at 4.27 $\micron$. The bottom panel shows the spectrum of Phoebe, combining the G235H and G395M exposures and binning down the G235H resolutions down by a factor of 3 to match the G395M resolution. Our spectrum matches results from Cassini VIMS with clear water ice and CO$_2$ absorptions \citep{Clark2005Natur, Fraser2018AJ}. The lack of evidence for water ice on the smaller Albiorix and Siarnaq distinguishes them from the water-rich Phoebe, indicating either the smaller irregular satellites were initially depleted of their water ice, or that post-capture evolution removed the ice.}
    \label{fig:mainfig}
\end{figure*}
\section{Observations}
\label{sec:obs}
The observations of the three largest irregular satellites of Saturn -- Phoebe, Siarnaq, and Albiorix -- were taken using JWST NIRSpec as part of JWST Cycle 2 Program \#3716 (PI: M. Brown). The smaller Siarnaq and Albiorix were obtained in PRISM mode, which observes in the 0.6-5.3 $\micron$ range, with an R $\sim$ 30-300 resolution. Phoebe's spectrum was taken using two different gratings, at high resolution in G235H/F170LP mode which covers 1.66–3.05 $\micron$ at an R $\sim$ 2700 resolution, and at medium resolution in G395M/F290LP mode which covers 2.87–5.10 $\micron$ at resolution of R $\sim$ 1000. The observations used the NRSIRS2RAPID readout method, which reads additional non-science reference pixels to reduce the correlated noise caused during detector readout. Each satellite's ephemerides were sufficiently accurate to blindly acquire the targets into the 3$\times$3 arcsecond aperture. Each integration was performed using a four-point dither to ameliorate the effects of bad pixels. Observational details are summarized in Table 1. 

The raw ``uncal.fits'' image cubes cubes are processed with version 1.15.1 of the official JWST calibration
pipeline and calibration files \citep{bushouse_2024_12692459}. We proceed with the standard Step 1 and 2 pipelines, with a custom noise removal routine performed after the rate step, as described in \cite{Brown2023PSJ}. The spectra are extracted from the Step 2 ``s3d.fits'' image cubes using a custom PSF-fitting algorithm that mitigates the wavelength-dependent features of the PSF, similarly to reductions described in \cite{Brown2023PSJ} and \cite{Wong2024PSJ}. First, the background is subtracted by taking the median of spatial pixels in a region outside a 10$\times$10 pixel box centered on the target. Then, at each wavelength slice of the data cube, we model the PSF by taking a two-dimensional median over a 7$\times$7 pixel area around the target across all wavelength slices within $\pm$ 50 wavelength bins in PRISM mode and $\pm$ 30 wavelength bins in the higher resolution mode, and normalizing the resulting template to sum to unity. The irradiance is then the best-fit parameter to the ratio of the data in each wavelength bin to the template empirical PSF, found by using a median weighted on the amplitude of the emperical PSF in each pixel. We then median-average the four dithers to obtain the final 1-D spectra, though we drop the final dither for Albiorix as there is a background source within the extraction aperture. The spectra are divided by a solar analog obtained from Program 1128 (PI: N. Luetzgendorf) to obtain the relative reflectance \citep{Gordon2022AJ}.

\section{Spectral features}
\label{sec:spec}
The relative reflectance spectra of Albiorix, Siarnaq, and Phoebe are shown in \autoref{fig:mainfig}. Upon inspection, Albiorix and Siarnaq look distinct from Phoebe in the 1.5 to 5.0 $\micron$ range. While Phoebe has significant water ice and a deep 3 $\micron$ O-H band, Albiorix and Siarnaq do not have apparent signatures of water ice as they do not have a Fresnel peak at 3.1 $\micron$, the 1.5 overtone (2$\nu_3$) or 2.0 $\micron$ combination feature ($\nu_1+\nu_3$), or the broad 4.5 $\micron$ feature ($\nu_2+\nu_r$) \citep{Mastrapa2008Icar,Mastrapa2009ApJ}. All three objects have visible absorptions from CO$_2$ at 4.27 $\micron$.

\begin{figure}
    \centering
    \includegraphics[width=8.6cm]{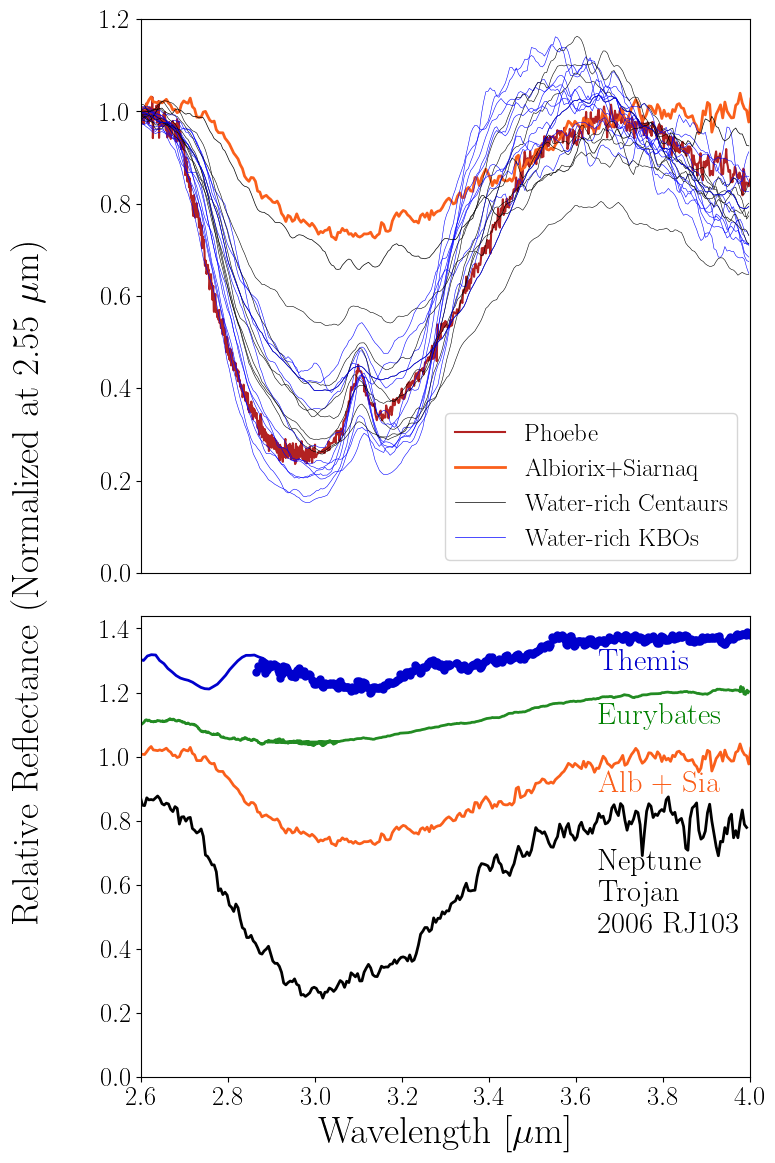}
    \caption{Top Panel: Comparison of the 3 $\micron$ feature on Phoebe (red), Siarnaq and Albiorix (orange), and water-rich Centaurs (grey) and KBOs (blue) from Program 2418 (PI: N. Pinilla-Alonso). Phoebe's 3 $\micron$ feature is similar to that of other water-rich KBO spectra, while Albiorix and Siarnaq are more similar to the most processed Centaurs, such as Okyrhoe. Bottom panel: Comparison of the 3.0 micron feature across representative small bodies, Eurybates (green, +0.3 offset, from \citealt{Wong2024PSJ}), Themis (blue, +0.15 offset, from \cite{Rivkin2010Natur} and \citealt{Usui2019PASJ}), Siarnaq and Albiorix (orange), and the Neptune Trojan 2006 RJ103 (black, -0.2 offset, from \citealt{Markwardt2023arXiv}). The Jupiter Trojans and asteroid belt objects are distinct from the Saturnian irregulars, with significantly weaker 3.1 $\micron$ absorptions. The Neptune Trojan 2006 RJ103 has a deeper 3.0 $\micron$ band than the small Saturnian irregulars, but also lacks water ice features at other wavelengths.}
    \label{fig:3micron}
\end{figure}

\subsection{O-H Stretch and Water Ice}
The broad 3 $\micron$ band associated with the O-H stretch is a common feature of the infrared spectra of outer solar system small bodies. The Jupiter Trojans observed with JWST \citep{Wong2024PSJ} show very shallow and broad 3 $\micron$ O-H bands that are distinct from those in the asteroid belt associated with hydrated silicates \citep{Takir2012Icar, Usui2019PASJ}, but are also much less deep than on Centaurs and KBOs, possibly a result of the significant thermal processing at 5.2 au which removes water ice from their surfaces \citep{Lepoutre2014Icar}. Centaurs and KBOs have a greater variety of 3 $\micron$ bands, with some indicating the presence of water ice through the presence of the Fresnel peak at 3.1 $\micron$, as well as features at 1.5 and 2.0 $\micron$ \citep{Pinilla-Alonso2024NatAs}.

In \autoref{fig:3micron} we show the 3 $\micron$ feature of the Saturnian irregular satellites against representative small body spectra: water ice-rich KBOs and Centaurs from JWST programs 1191 (PI: J. Stansberry) and 2418 (PI: N. Pinilla-Alonso) reduced using the pipeline described here, the main belt Themis from \cite{Rivkin2010Natur} and \cite{Usui2019PASJ}, the Jupiter Trojan Eurybates from \cite{Wong2024PSJ}, and the Neptune Trojan 2006 RJ103 from \cite{Markwardt2023arXiv}. Phoebe's 3 $\micron$ feature has a 75\% band depth and a sharp Fresnel peak at 3.1 $\micron$, similar to most water-rich KBOs. Phoebe also shows a downwards slope in the spectrum past 3.6 $\mu$m, attributable to the broad 4.5 $\micron$ combination band seen in \autoref{fig:mainfig} \citep{Mastrapa2009ApJ}. By contrast, Albiorix and Siarnaq have a 3 $\mu$m feature with 25\% band depth, and no evidence for either the Fresnel peak or the 4.5 $\mu$m combination band. Phoebe is similar to the most water-rich KBOs, while Albiorix and Siarnaq most resemble centaurs such as Okyrhoe and 2013 XZ8 which have muted 3 $\mu$m features \citep{Licandro2024NatAs}. The two small Saturnian irregular satellites are both distinct from the Jupiter Trojans or main belt asteroids, owing to their much larger 3 $\micron$ band depths.

\begin{figure}
    \centering
    \includegraphics[width=8.5cm]{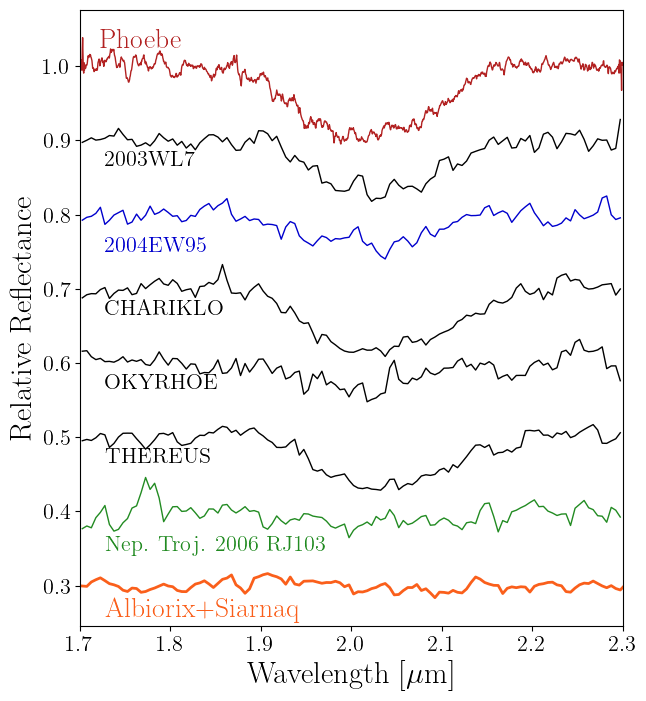}
    \caption{Comparison of the continuum-removed 2.0 $\micron$ feature on Phoebe (red), Siarnaq and Albiorix (orange), selected Centaurs and the Plutno 2004 EW95 from \citealt{Licandro2024NatAs, Pinilla-Alonso2024NatAs}, and the Neptune Trojan 2006 RJ103 from \citealt{Markwardt2023arXiv}. Each spectrum below Phoebe is successively offset by 0.1 down in relative reflectance. All of the KBOs and Phoebe show 2.0 $\micron$ absorptions, ranging from 3.5\% band depth on Okyrhoe to 9\% on Chariklo. By contrast, Siarnaq and Albiorix, as well as 2006 RJ103, show no evidence for a 2.0 $\micron$ band, suggesting a lack of exposed water ice.}
    \label{fig:2micron}
\end{figure}

To confirm that the smaller irregular satellites lack pure water ice on their surface, we also look at the 1.5 $\mu$m water ice overtone and 2.0 $\mu$m water ice combination band. For the 2.0 $\micron$ feature, we find a best-fit linear continuum between 1.7 and 2.3 $\micron$, ignoring the band between 1.9 and 2.15 $\micron$, then fit a simple Gaussian to the continuum-subtracted band. In \autoref{fig:2micron}, we show the continuum-subtracted 2.0 $\micron$ feature on Phoebe, the two smaller irregular satellites, and five KBOs and Centaurs. Siarnaq and Albiorix show no evidence for the 2.0 $\micron$ band. We find a 1.0\% upper limit on the feature's band depth for Albiorix and Siarnaq, which we estimate by first determining the noise level surrounding the band, then adding Gaussian signal with the same band width and center as on the other KBOs until our fitting routine detects a band depth that is 3$\sigma$ above the noise level. We avoid translating an upper limit on the band depth to a fraction of water ice as modeling using the available 1-5 $\micron$ range will produce degeneracies between grain size and abundance. While the differences in the appearance of the 3 $\micron$ feature between the KBOs, Centaurs, and Trojans could plausibly be explained by their different thermal histories, Albiorix, Siarnaq, and Phoebe have experienced the same level of solar insolation since capture, so some other mechanism must have led to the difference between them. The removal of water ice from the smaller irregular satellites is either a result of some other feature of post-capture evolution such as collisions, or a primordial lack of water ice for the smaller objects. We discuss these possible explanations for the lack of water ice on the smaller irregular satellites in \autoref{sec:discussion}.

\subsubsection{Compositional attribution of the 3 $\micron$ band}
The distinct shape of the 3 $\micron$ feature on Albiorix and Siarnaq combined with the lack of other typical signatures of water ice in the 1-5$\micron$ range makes suggesting a plausible composition challenging. The main belt asteroid Themis, whose spectrum is shown in the bottom panel of \autoref{fig:3micron}, has a 3.0 $\micron$ band without the typical 1.5 or 2.0 $\micron$ features. The 3.0 $\micron$ feature on Themis has been attributed to silicate grains coated with sub-micron sized water ice grains \citep{Rivkin2010Natur}, though this model has received several observational challenges in the form of non-detections of water-ice sublimation \citep{Jewitt2012AJ, McKay2017Icar, O'Rourke2020ApJL}.The band depth and shape of the feature is significantly different on Siarnaq and Albiorix than on Themis, and Themis has a 2.7 $\micron$ phyllosilicate band which does not appear on Siarnaq or Albiorix \citep{Usui2019PASJ}. Moreover, replicating the Shkuratov model of \cite{Rivkin2010Natur} or \cite{Brown2016AJ} for the small silicate grains covered in water but at a higher abundance does not reproduce the shape of the broad feature seen on the smaller Saturnian irregular satellites. We therefore argue against significant quantities of water ice on Siarnaq and Albiorix.

A hint to a possible compositional attribution of the 3 $\micron$ band on Albiorix and Siarnaq lies in their resemblance to the Neptune Trojan 2006 RJ103, which also does not show evidence for significant surface water ice \citep{Markwardt2023arXiv}. Ground-based visible spectroscopy of 2006 RJ103 shows a strong 600-700 nm feature associated with the Fe$^{2+}$/Fe$^{3+}$ charge transfer \citep{Sharkey2023PSJ}, which combined with its 3.0 $\micron$ O-H feature shown in \autoref{fig:3micron} is indicative of the presence of iron-bearing hydrated minerals. It is therefore plausible that the deep 3.0 $\micron$ bands on Albiorix and Siarnaq (relative to the Jupiter Trojans and main belt objects) are a consequence of abundant hydrated minerals. Phoebe also has a clear 1.0 $\micron$ absorption \citep{Clark2005Natur, Coradini2008Icar} that is deepest in the least water-rich areas \citep{Fraser2018AJ}, attributed to some unidentified iron-bearing mineral, possibly hydrated \citep{Clark2005Natur, Clark2012Icar}. While we do not see an identical feature on Albiorix and Siarnaq, there is a slope change at 1.15 $\micron$ in \autoref{fig:mainfig}, consistent with an absorber in the 0.7-1.0 $\micron$ range. One issue with invoking hydrated minerals is that most show a narrow 2.7 $\micron$ absorption, such as the one on Themis in \autoref{fig:3micron}. However, there are hydrated minerals and meteorite samples without this band, such as the Murchison meteorite and the minerals cronstedtite and goethite \citep{Bates2020M&PS}; goethite has been suggested to be responsible for the 3.1 $\micron$ feature in main belt asteroids \citep{Beck2011A&A}.

\subsection{Organic material}
The albedos of Phoebe, Siarnaq, and Albiorix are significantly darker than those of the regular Saturnian satellites and are more similar to those typical of Kuiper belt objects \citep{Lellouch2013A&A}. The dark material from Phoebe coats Hyperion and the leading side of Iapetus and is an important feature of the surface material in the Saturnian system \citep{Cruikshank1983Icar, Verbiscer2009Natur, Clark2012Icar, Cruikshank2014Icar}, but a specific identification of the composition of the dark material remains elusive. We examine the 3.2-3.6 micron region of the irregular satellite spectra for potential absorptions by the C-H stretch typical of aromatic and aliphatic hydrocarbons.

\begin{figure*}
    \centering
    \includegraphics[width=0.9\textwidth]{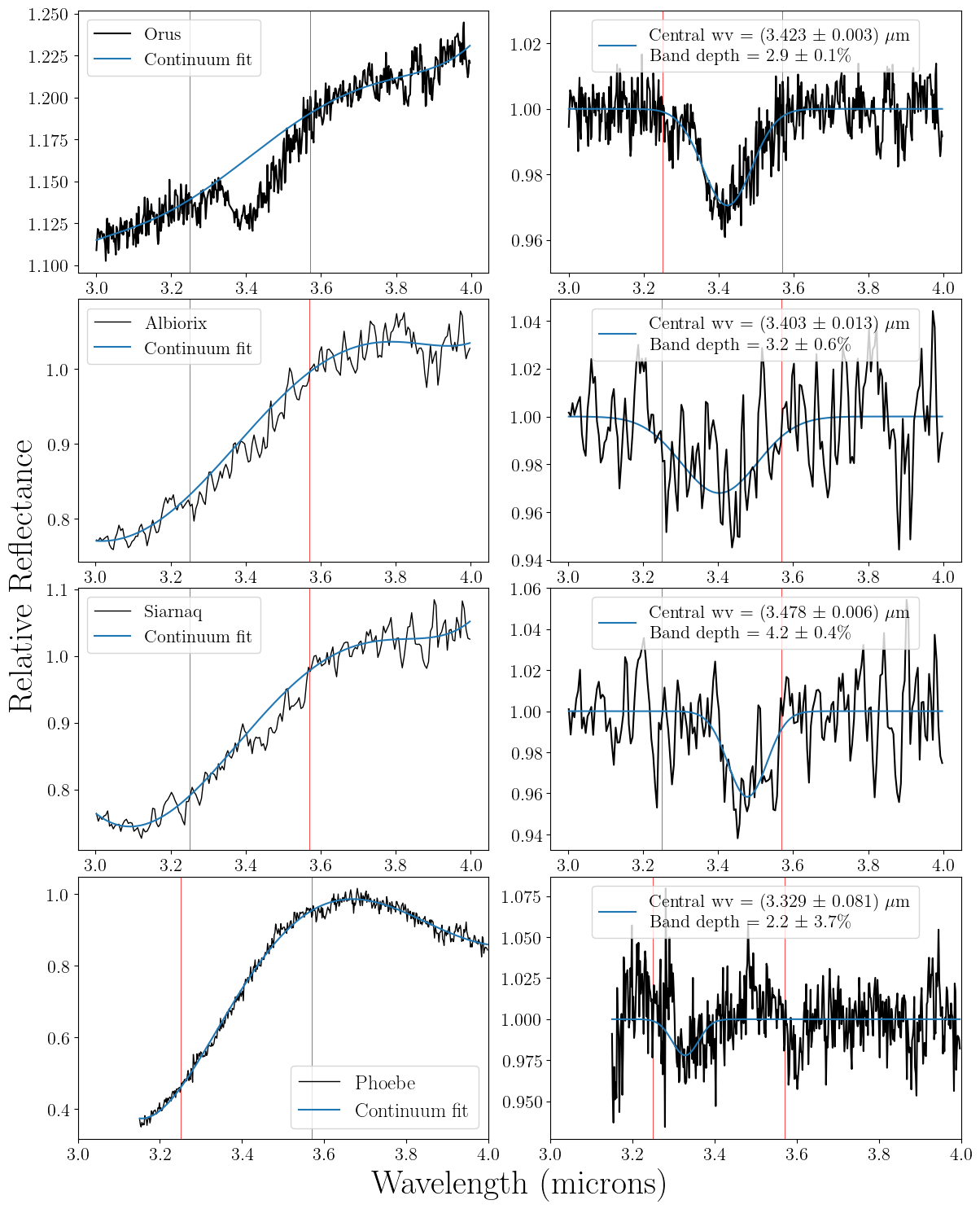}
    \caption{Left panels show the 3-4 $\micron$ spectra and polynomial continuum fits of Orus, Albiorix, Siarnaq, and Phoebe (from top to bottom). The red lines show the area ignored by the continuum fit, which is the same area used by \cite{Wong2024PSJ}. Right panels show the continuum-subtracted spectrum, with Gaussian fits to the organic band. Both individually and in their average, the small Saturnian irregular satellites show organic bands that are consistent with what is seen on the Trojans, but at slightly longer wavelength. Phoebe does not show strong evidence for an absorption in the 3.25-3.6 $\micron$, obscured by the water ice 3 $\micron$ feature which makes determining a continuum difficult.}
    \label{fig:organics}
\end{figure*}

One common explanation for the low albedos of outer solar system small bodies is the presence of an organic darkening agent \citep{Cruikshank1998Icar,Khare2001AdSpR, Barucci2011Icar}. In the infrared, organic compounds and their irradiation products have absorption features through the 3-4 $\micron$ region that are observed throughout the outer solar system: large KBOs like Eris or Pluto have simpler organics \citep{Grundy2016Sci, Emery2024Icar}, comets show emission features attributed to aliphatic compounds \citep{Raponi2020NatAs, Kelley2023Natur}, and KBOs have broad 3.4-3.6 $\mu$m absorptions (Figure 1 of \citealt{Brown2023PSJ}). Irradiation and sublimation destroy and remove the simplest organics from the surface, with the terminal stage of such processing resulting in amorphous, largely dehydrogenated, carbon \citep{Moore1998Icar, Moroz2004Icar, Clark2009JGRE, Hand2012JGRE}. We expect the Saturnian irregular satellites to show organic bands similar to the Jupiter Trojans, which have broad 3.4 $\mu$m absorptions.

To characterize the organics on the irregular satellites we isolate the 3.3-3.6 $\micron$ area that could potentially host organic absorptions and try to see if there is an excess absorption, following the method laid out in \cite{Wong2024PSJ}. We perform a continuum fit to the area between 3.0 and 4.0 $\micron$ using a fifth degree polynomial, excluding the 3.25-3.57 $\micron$ area that potentially has organic absorption features. Repeating the exercise with a third-degree polynomial using the same region or a linear fit over a smaller range results in a qualitatively similar answer, with a less consistent result for the center of the band on Albiorix. A shorter wavelength range is used for Phoebe as the water feature Fresnel peak at 3.1 $\micron$ impedes the continuum fit. In \autoref{fig:organics} we show the continuum fits to the Jupiter Trojan Orus, Albiorix, Siarnaq, and Phoebe, and their average in the left set of panels, and the continuum-subtracted feature with Gaussian band fits in the right set of panels. We use a Monte Carlo procedure for fitting the band. Using the errors on each data point, we resample the spectra into 10,000 simulated spectra. To each continuum-subtracted spectrum, we fit a Gaussian using a least squares algorithm. The means and standard deviations of the resulting parameter distributions are the reported values and uncertainties for the band centers and depths in \autoref{fig:organics}. 

The parameters of the 3.4$\mu$m feature on the smaller Saturnian irregular satellites are similar to Orus, which, of the five Trojans observed in \cite{Wong2024PSJ}, has the deepest 3.4$\mu$m absorption. The organic absorption on Albiorix has a band center and band depth within the margin of error of the fits for Orus from both our work and \cite{Wong2024PSJ}. Siarnaq has a slightly longer-wavelength and narrower band, though by eye the absorption is visible as a plateau at 3.45 $\micron$ even without continuum subtraction. The organic band at 3.4 $\mu$m we observe on the small irregular satellites is consistent with the C-H stretch, but the position is sensitive to the parent molecule the C-H group is attached to (c.f. Tables 112 and 115 in \citealt{Herzberg1945msms}). Laboratory experiments by \cite{Hand2012JGRE} show that the asymmetric mode of the -CH$_2$- band and the asymmetric and symmetric modes of the -CH$_3$- band in 70–100 K water ice/long-chain aliphatic mixtures are centered in the 3.38-3.49 area that our bands cover, and are stable not only to irradiation, but also to warming to 300 K. 

Phoebe shows a potential absorption at 3.3 $\micron$ that is short of the other irregular satellites. Whether the feature appears is extremely sensitive to the choice of anchor points for the continuum subtraction and the subset of dithers used to build the spectrum. In addition to the analysis we present in \autoref{fig:organics}, we repeat the spline continuum removal done by \cite{DalleOre2012Icar} and \cite{Cruikshank2014Icar} for the \textit{Cassini} VIMS data of Phoebe on our data, but do not find evidence for any of the same aliphatic organic absorptions previously claimed on Phoebe. To fully extract the signatures of organic compounds, continuum removal is insufficient, as the water ice feature has complex behavior in the 3.0-3.6 $\micron$ range with subtle changes in slope at 3.3-3.5 $\micron$ that obscure the signatures of organic molecules.


\subsection{4.27 $\mu$m CO$_2$ Feature}

JWST spectroscopy has demonstrated that CO$_2$ is common on outer solar system small bodies, even on those that experience temperatures above 80K, at which pure CO$_2$ ice sublimates \citep{Brown2023PSJ,Wong2024PSJ, DePra2024NatAs}. On KBO surfaces, the CO$_2$ band has a wide range of appearances. The spectra of the most distant and organic-rich objects show evidence for Mie scattering caused by micron thick layers of small CO$_2$ grains on their surfaces \citep{Brown2023PSJ}. By contrast, the water-rich KBOs and Centaurs have what is likely CO$_2$ trapped in the water ice, organics, or silicates that are possibly present on their surfaces \citep{DePra2024NatAs}. The band centers of the CO$_2$ features range between 4.250-4.275 $\micron$, with the water-ice rich KBOs tending to have longer-wavelength band centers than the organic-rich KBOs (see figure 4 of \citealt{DePra2024NatAs}). On the Jupiter Trojans observed thus far, only Eurybates has shown evidence for a CO$_2$ feature, with a band center of 4.258 $\pm$ 0.002, more similar to the central wavelength of the organic rich KBOs than the water-rich ones \citep{Wong2024PSJ}. 

CO$_2$ was previously detected in the Cassini VIMS spectrum of Phoebe \citep{Clark2005Natur}, with \cite{Hansen2012Icar} finding that the spatial distribution of the CO$_2$ on Phoebe has a slight inverse correlation with water ice and has a globally averaged band depth of approximately 20\%. Improved abundance maps of water ice from \cite{Fraser2018AJ} also suggest the same inverse correlation between CO$_2$ and surface water ice (comparing the maps in Fig. 4 of \citealt{Fraser2018AJ} and Fig. 10 of \citealt{Hansen2012Icar}). Further analysis by \cite{Clark2019Icar} also claims a detection of $^{13}$CO$_2$ on Phoebe, at 5\% of the equivalent width or 10\% of the band depth of the 4.27 CO$_2$ feature.

\begin{figure}
    \centering
    \includegraphics[width=9cm]{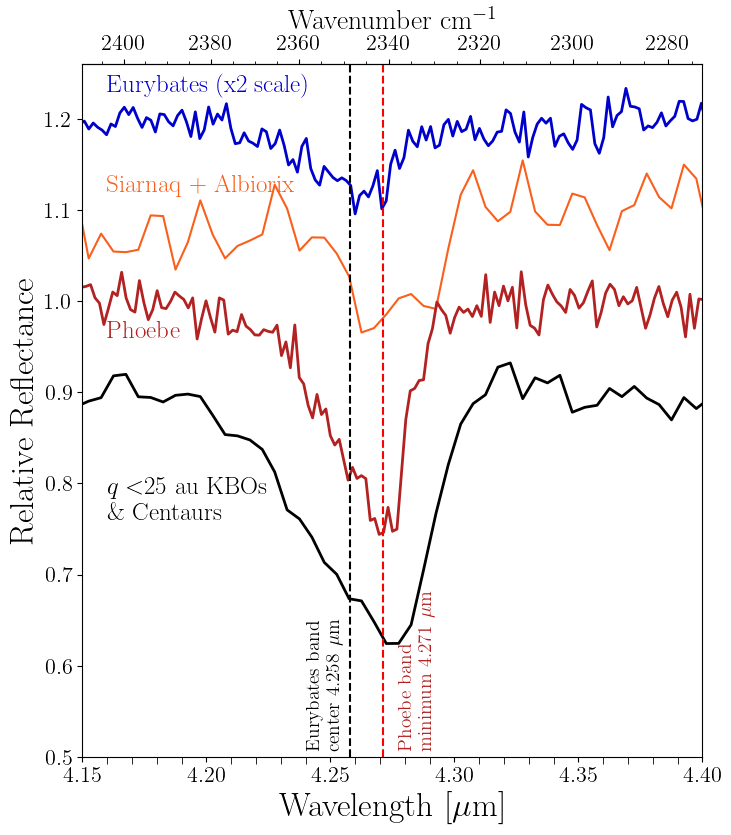}
    \caption{Continuum-removed CO$_2$ feature on, from top to bottom, Jupiter Trojan Eurybates, Albiorix and Siarnaq, Phoebe, and the average spectrum of $q<25$ au KBOs and Centaurs. The band center of the CO$_2$ feature on Phoebe is similar to that of the Centaurs, indicating a common trapping mechanism, as pure CO$_2$ ice is not stable at Saturnian temperatures. Both Phoebe and the KBOs have a shallower short-wavelength slope that is not present on the CO$_2$ on the smaller irregular satellites. CO$_2$ on Eurybates has a distinct 4.258 $\mu$m band center \citep{Wong2024PSJ}, possibly indicating that a different medium is trapping the CO$_2$ on Eurybates or that the wavelengths are temperature
    dependent.}
    \label{fig:co2}
\end{figure}

In \autoref{fig:co2}, we show the continuum removed CO$_2$ bands of Phoebe, Siarnaq and Albiorix averaged, Eurybates, and the median of all $q<25$ au KBOs and Centaurs. We remove the continuum by fitting a line between 4.0 and 4.5 $\mu$m, excluding points between 4.20 and 4.33 $\mu$m. We then repeat the fitting procedure described above. Fitting a single Gaussian to the CO$_2$ band of the small irregular satellites, we find that Siarnaq has a CO$_2$ band center of 4.272 $\pm$ 0.002 $\mu$m, Albiorix's band is at 4.275 $\pm$ 0.004 $\mu$m, their average has a band center of 4.272 $\pm$ 0.001 $\mu$m and a band depth of 13.1 $\pm$ 0.6\%. Phoebe's band minimum is at 4.271 $\mu$m, but the feature appears to be best modeled by the sum of two Gaussians, centered at 4.258 $\mu$m and 4.272 $\mu$m respectively. The Centaur median CO$_2$ band is similar to Phoebe, with a similar, though slightly longer band minimum, suggesting a common mechanism for formation and trapping of CO$_2$ inside its sublimation line on KBO surfaces. Contrary to \cite{Clark2019Icar}, we do not see evidence above the local noise level for the $^{13}$CO$_2$ feature on Phoebe (noise at the wings of the feature is 1.7\%), ruling out the \cite{Clark2019Icar} claimed $0.053$ band depth at 3 $\sigma$. Recent work has shown that the $^{13}$C/$^{12}$C ratio in the methane on Eris and Makemake, as well as comet 67/P is similar to terrestrial values \citep{Muller2022A&A, Grundy2024Icar}, and individual water-rich KBOs do not show clear evidence of the 4.36 $\micron$ feature \citep{DePra2024NatAs}. The absence of the $^{13}$CO$_2$ feature in the JWST spectrum of Phoebe combined with a set of similar results for the isotopic ratios of other bodies of outer solar system origin suggests that the $^{13}$C/$^{12}$C ratio on Phoebe is likely more similar to that of Iapetus or terrestrial values, than the one determined in \cite{Clark2019Icar}. 

CO$_2$ on the Saturnian irregular satellites is at temperatures well above its sublimation temperature, raising the question of what is trapping the CO$_2$.  Brown et al. (2024, submitted) find that on all the outer Saturnian satellites (Hyperion, Iapetus, and Phoebe), the 4.271 $\mu$m CO$_2$ absorption is linked to the dark material derived from Phoebe. Organic compounds are a common explanation invoked for the low-albedo material both on Phoebe and on KBOs in general \citep{Cruikshank1998Icar, Khare2001AdSpR,Brown2012AREPS, Cruikshank2014Icar}. Irradiation of organics such of asphaltite layered with water ice in laboratory experiments from \cite{Strazzulla2005A&A} and \cite{Gomis2005Icar} result in a 4.270-4.272 $\micron$ feature consistent with the band center of CO$_2$ on all three Saturnian irregular satellites. Albiorix and Siarnaq have evidence for a C-H feature at 3.4 $\mu$m, which strengthens the claim that organics are the source of the CO$_2$. Phoebe lacks an obvious detection of organics, but the iron oxides detected by Cassini VIMS \citep{Clark2012Icar} could also efficiently trap CO$_2$ \citep{F19757101623}.

\section{Discussion} \label{sec:discussion}
The three Saturnian irregular satellites have spectra that are most similar to those of the water-rich Centaurs and KBOs observed thus far, \citep{Licandro2024NatAs,Pinilla-Alonso2024NatAs}, as is expected from the dynamical picture of irregular satellite capture. Less clear is why Albiorix and Siarnaq lack water ice, while Phoebe has ubiquitous surface water ice \citep{Clark2005Natur, Fraser2018AJ}. 

We hypothesize that the smaller irregular satellites once had water ice like Phoebe and many of the Centaurs, but that this water ice has been removed by sublimation during catastrophic collisions. This mechanism is a unique consequence of the specific environment of the irregular satellite cloud. Unlike the Kuiper belt, collisional velocities among irregular satellites are very high -- collisions between a prograde and retrograde object can reach 5 km/s relative speeds. Moreover, irregular satellites have notably high collisional probabilities post-capture, at least seven orders of magnitude higher than in the Kuiper belt (compare \citealt{Bottke2010AJ} and \citealt{Abedin2021AJ}). 

We can compare the energy of an impact between irregular satellites to the latent heat of ice, which is a first-order check for plausibility of collisional removal of the water ice. Assuming a temperature of 110K, and using the temperature-dependent expression for the latent heat of vaporization of water ice from \cite{2005QJRMS.131.1539M}, we find that the water ice will have a heat of vaporization of just under 2.8 MJ/kg, while if we assume Siarnaq and Albiorix have a head-on collision near perihelion (a combined relative velocity of 5 km/s) with equal mass, their collision will have a $v_{imp}^2/8=$ 3.1 MJ/kg specific impact energy over the total mass of the two bodies. Even though much of the impact energy will go to excavation and removal of mass to infinity rather than thermal processes, significant vaporization seems likely. An additional argument in favor of the water being fully processed by collisions is the high pressure at the point of contact, which is approximately $\rho \times v_{imp}^2$ \citep{Melosh1989icgp, Barr2016MNRAS}, resulting in significantly higher than 1 GPa pressures for these catastrophic collisions, more than enough to melt significant amounts of water ice. Shock wave experiments by \cite{stewart_ahrens_2005} show that at 100K, bodies impacting with 4 GPa pressures will be completely melted. Simulations from \cite{Jutzi2020Icar} predict that for two equal-sized bodies, a 5 km/s relative impact velocity will result in enough of a temperature increase to melt all water ice present, and sublimate around 10\% of the ice. This melting would leave behind hydrated minerals that are a likely explanation for the deep O-H feature on the smaller irregular satellites as compared to the Jupiter Trojans.

\begin{figure}
    \centering
    \includegraphics[width=8.5cm]{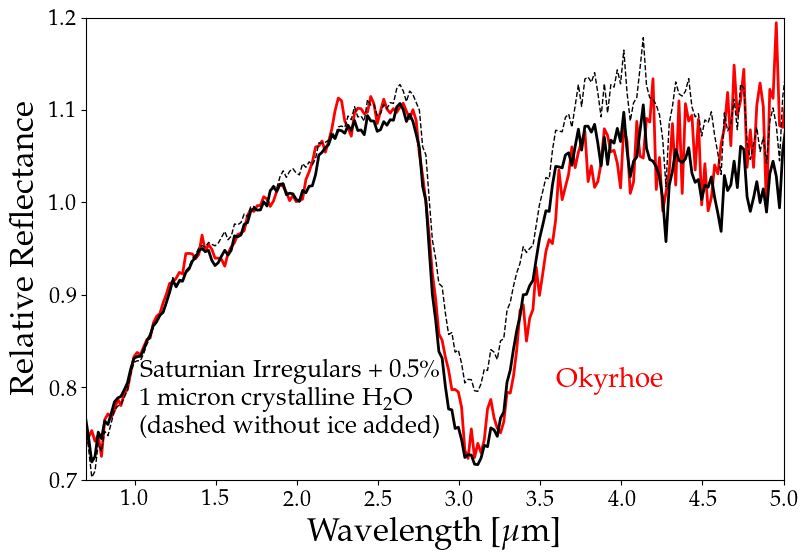}
    \caption{Average of Siarnaq and Albiorix spectra with a layer of 10 $\micron$ grain size crystalline water ice added using Hapke modelling (original dashed, with water in black). Adding water ice as a singular component to the irregular satellites in an areal mixture results in a spectrum that matches Okyrhoe (red), indicating the irregular satellites are consistent with having once had a water-ice rich spectrum.}
    \label{fig:watermodel}
\end{figure}

We can visualize how the spectra of Siarnaq and Albiorix originally might have appeared by combining their spectra with a modeled water ice spectrum. We model the ice using Hapke theory from \cite{Hapke1981JGR, Hapke1993tres} and optical constants for water ice from \cite{Clark2012Icar, Clark2019Icar}. We use the Python package \texttt{miepython} \citep{prahl_2023_7949403} to find the single-scattering albedo of a spherical water particle of a given diameter using the \cite{Clark2019Icar} optical constants. Then, using Equation (9.9) of \cite{Hapke1993tres} we obtain the diffusive reflectance which we then convert to geometric albedo using the approximation given in Equation (10.39). By adding 10 micron water ice grains at a 0.4 \% abundance in a linear combination with the original spectrum of the small irregulars, we obtain a spectrum for the small irregular satellites with surface patches of water ice. An areal mixture is sensible in this case, as collisions are known to expose water ice on Phoebe \citep{Fraser2018AJ}.This spectrum, shown in \autoref{fig:watermodel}, closely matches the spectrum of the Centaur Okyrhoe, demonstrating the spectral similarity of the non-ice component of these objects. Reversing this operation, if we were to remove the water ice signature from Okyrhoe, its spectrum would resemble that of the small Saturnian irregulars. If water ice were present on Albiorix and Siarnaq, collisions would expose it in patches, which would then match the model in \autoref{fig:watermodel}. As a Centaur, Okyrhoe has not experienced the significant collisional history of an irregular satellite yet has retained its water ice, consistent with our hypothesis of collisions then removing water ice from Albiorix and Siarnaq.

Unlike the smaller irregular satellites, Phoebe retains its water ice. At 200km in diameter, Phoebe is significantly larger than that of any other Saturnian irregular satellite, and plausibly an intact planetesimal. By contrast, Siarnaq and Albiorix are at $\sim$40 km in diameter and parent members of the Gallic and Inuit satellite families, indicating past disruption of a parent body \citep{Denk2018eims}. Images of Phoebe show a highly-cratered surface with the largest craters nearing the catastrophic disruption limit \citep{Porco2005Sci, Bottke2010AJ}. Impactors that cause craters of that size would fully disrupt the smaller Albiorix and Siarnaq. Disrupting Phoebe would require a body of at least 20 km diameter (computed using equation 4 and figure 7 of \citealt{Bottke2010AJ}), and an order of magnitude fewer such objects are captured than the 4 km diameter objects required to disrupt Siarnaq or Albiorix \citep{2007AJ....133.1962N}. Phoebe retains its water ice because it has not experienced a catastrophic collision, as evidenced by its near-spherical shape, high density, low bulk porosity, and evidence of structure and geology from images \citep{Porco2005Sci, Johnson2005Natur, Jacobson2006AJ, Thomas2010Icar}. Melt from the collisions on Phoebe possibly accumulates at crater bottoms, explaining the observed higher abundance of water ice at the bottom of craters \citep{Fraser2018AJ}. 

We posit that post-capture sublimation is not a plausible explanation for the differences between Phoebe and the smaller irregular satellites, as they have been at the same heliocentric distance since capture. On Phoebe, even the darkest locations have water ice \citep{Fraser2018AJ} and have 3.0 $\micron$ band depths significantly exceeding Albiorix and Siarnaq. Finally, water ice is stable at temperatures in the Saturnian system. While Jeans escape will deplete some amount of water ice from the surface, even the low albedo leading side of Iapetus and dark parts of Phoebe have clear signs of water ice.

An alternative explanation for the spectra of the Saturnian irregular satellites is a different natal environment for Phoebe than Siarnaq and Albiorix, which would presumably then be formed in a less water-rich region of the protoplanetary disk. Our comparison of Siarnaq and Albiorix to Okyrhoe suggests that a distinct natal environment for observed water-poor bodies is a less plausible explanation than collisional processing, but to completely disfavor this hypothesis would require observation of a significant sample of similar-sized KBOs ($H\sim11$) to demonstrate they retain their water ice. 

The final piece of the spectroscopic puzzle is the presence of CO$_2$ on the smaller irregular satellites despite their potentially extreme past thermal and collisional history. The implication here is that CO$_2$ is formed in-situ from solar irradiation of available materials. As we demonstrate in \autoref{fig:organics}, organics at $3.4\micron$ are a possible component of the spectra of the smaller irregular satellites, providing the necessary material to form CO$_2$. The presence of CO$_2$ on these heavily processed bodies suggests that radiolytic formation and subsequent trapping of CO$_2$ is a common process in the outer solar system when organics are found in conjunction with an O-H bearing material that can supply the oxygen necessary to form CO$_2$, such as the one causing the 3 micron band on the smaller irregular satellites. 

\subsection{The Trojan-Irregular-KBO Connection}
While the spectral similarity of Phoebe to water-rich KBOs argues strongly in favor of a common origin for the irregular satellites and present-day Kuiper belt objects, there remains a tension between the  difference between the observed distribution of visible colors in the redder Kuiper belt and the bluer captured populations: the giant planet irregular satellites and the Jupiter and Neptune Trojans \citep{Wong2016AJ, Graykowski2018AJ, Jewitt2018AJ, Bolin2023MNRAS, Sharkey2023PSJ}. For Jupiter's Trojans and Irregulars, thermal insolation is commonly invoked to explain the change from Kuiper belt colors to the present-day relatively neutral visible colors \citep{Wong2016AJ, Sharkey2023PSJ}. The depletion of water from the smaller Saturnian irregular satellites is evidence that collisions can induce significant change in the surface properties of irregular satellites at large, which could potentially explain why at Saturn and Uranus the irregular satellites are notably bluer than the Kuiper belt.  Finally, JWST spectroscopy of the Neptune Trojans by \cite{Markwardt2023arXiv} found that the Neptune Trojans span the same range of spectral types as the Kuiper belt objects presented in \cite{Pinilla-Alonso2024NatAs}, though perhaps in different proportions of organic-rich to water-rich types as evidenced by statistics on their visible coors \citep{Bolin2023MNRAS}. Thus far, the published JWST NIRSpec small body data suggests that each of the captured populations linked by dynamical instability theories to the Kuiper belt is either spectrally analogous to the larger Kuiper belt population or one of its subsets, or has a population-specific explanation for its difference from observed Kuiper belt object spectra.

\section{Acknowledgments}
This work is based on observations made with the NASA/ESA/CSA James Webb Space Telescope. The data were obtained from the Mikulski Archive for Space Telescopes (MAST) at the Space Telescope Science Institute, which is operated by the Association of Universities for Research in Astronomy, Inc., under NASA contract NAS 5-03127 for JWST. These observations are associated with program \#3716. Support for program \#3716 was provided by NASA through a grant from the Space Telescope Science Institute which is operate by the Association of Universities for Research in Astronomy, Inc., under NASA contract NAS 5-03127. The specific observations analyzed can be accessed via DOI: \dataset[10.17909/y22z-yt33]{https://doi.org/10.17909/y22z-yt33}.  The authors thank Dave Stevenson, Ryleigh Davis, and Morgan Saidel for useful discussions.

\facility{JWST (NIRSpec)}

\software{Astropy \citep{2022ApJ...935..167A}, \texttt{lmfit} \citep{newville_2015_11813}, JWST pipeline version 1.15.1 \citep{bushouse_2024_12692459}.}  

\bibliography{refs}{}
\bibliographystyle{aasjournal}

\end{document}